%Paper: hep-th/9206072
%From: RCM@hep.physics.mcgill.ca
%Date: Thu, 18 Jun 1992 14:23:33 -0400 (EDT)
%Date (revised): Tue, 23 Jun 1992 13:30:57 -0400 (EDT)

\input jnl.tex

\def\part{\partial}
\def\veps{\varepsilon}
\def\eps{\epsilon}
\def\ie{{\it i.e.,}\ }
\def\cit#1{[\cite{#1}]}
\def\hB{\hat{B}}
\def\part{\partial}
\def\parb{\bar{\partial}}

\def\partlrarrow{\setbox2=\hbox{$\scriptstyle\leftrightarrow$}%
\setbox1=\hbox to \wd2{\hfil$\partial$\hfil}%
\copy1\kern-\wd2\raise1.1\ht1\copy2}
\def\nablrarrow{\setbox2=\hbox{$\scriptstyle\leftrightarrow$}%
\setbox1=\hbox to \wd2{\hfil$\nabla$\hfil}%
\copy1\kern-\wd2\raise1.1\ht1\copy2}

\def\ssc{\scriptscriptstyle}
\def\scc{\scriptstyle}
\def\sla #1{\kern+.2em\raise.2ex\hbox{/}\kern-.65em{#1}}
\def\slap{\kern+.05em\raise.2ex\hbox{/}\kern-.5em\partial}
\def\-#1{_{\ssc {#1} }}
\def\reff#1{Reference [\cite{#1}]}
\def\s #1{{\cal {#1}}}
\def\ie{{\it i.e.,\ }}

\def\hat #1{\mathaccent94{#1}}
% paper
\rightline{McGill/92--21}
\rightline{hep-th/9206072}

\title WORMHOLES AND SUPERSYMMETRY
\author{J.R. Anglin\footnote{$^\dagger$}{anglin@physics.mcgill.ca} %
and R.C. Myers\footnote{$^\ddagger$}{rcm@physics.mcgill.ca}}
\affil  Physics Department, McGill University
Ernest Rutherford Building
3600 University Street
Montreal, Quebec, CANADA H3A 2T8
\abstract We derive an
effective local operator produced by certain wormhole instantons in a theory
containing a massless Wess-Zumino multiplet coupled to N=1 supergravity.
The induced interactions are $D$ terms, and hence will not lead
to spontaneous supersymmetry breaking. We
conclude that supersymmetry suppresses wormhole-induced matter couplings.

\endtitlepage

\eject

\head{1.  Introduction}

Tunneling amplitudes for spatial topology change in Euclidean quantum gravity
have become sources of
interest, speculation, and controversy.  The simple case of the ``wormhole"
instanton (a Euclidean metric configuration in which two asymptotically flat
regions are connected by a narrow ``throat") has been used to describe
tunneling processes in which baby universes are created and annihilated.  These
calculations typically show that wormholes lead in the low energy limit to
effective local interactions among matter fields[\cite{andyrev}].
In the present paper, we consider a wormhole in a theory of
a massless supersymmetric scalar multiplet
coupled to N=1 supergravity.  We find no evidence of wormhole induced
supersymmetry breaking in this model. We do find that, as long
predicted[\cite{k}],
supersymmetry cancels the simple scalar self-coupling found in
the comparable purely bosonic theory\cit{cole,ab}.

In the remaining part of this introduction, we will describe the basic features
of the argument we will follow.  In section 2, we consider a wormhole in
a bosonic theory, with a massless complex field coupled to Einstein gravity.
The derivation of the local operator induced by the wormhole is essentially
the same as that originally presented by \reff{cole}.
In section 3, we will introduce
supersymmetry, and obtain an original result exactly like that of section 2,
but in superspace; on this excuse we will refer to the instanton considered in
section 3 as a ``superwormhole."  Section 4 is a brief conclusion.  An
appendix follows,
containing the detailed supergravity calculations supporting section 3.

Our goal is to determine the effect at experimentally accessible scales of
wormholes that are very much smaller than those experimental
scales, yet sufficiently larger
than the Planck scale for us to use general relativity (and ultimately
supergravity) in our action.   At present, there is plenty of room
between those scales ---
 about 17 orders of magnitude --- within which to fit our wormholes.
Following Coleman and Lee\cit{cole}, we will construct
a field configuration in Euclidean four-space  in which a three-ball of radius
$r_{\ssc{0}}$ is cut out of a flat background space and replaced with
the end of a
wormhole, matching our fields at the boundary.  We will perform a saddle-point
approximation to the path integral using this ``cut and patch" configuration,
and we will find that the leading contributions to the action coming from the
wormhole insertion are boundary terms on the sphere at $r_{\ssc{0}}$.
Choosing $r_{\ssc{0}}$ to be small on the laboratory scale, these terms may
replaced by point-like interactions. Thus we arrive at a set of local
interactions induced by the wormholes in the effective low energy theory.

Note that wormholes in supersymmetric theories have
also been considered in References \cit{10,fay,gerry}.

\head{2.  The Bosonic Wormhole}
As a preliminary exercise, we consider wormholes in the purely bosonic theory
of a massless, complex scalar coupled to Einstein gravity.  The
essential results have been found previously in [\cite{cole, ab}]. Our analysis
differs slightly from the previous derivations, and we also find the
next of the
higher dimension operators induced by the wormholes.  While these
next-to-leading order
local terms play an insignificant role in the low energy bosonic
theory, they are important for the supersymmetric wormhole considered in the
following section.
In Euclidean space, we consider the Lagrangian  $$ \eqalign{ {\cal L}\-0&=\;
-e\bigl (M\-P ^2R\;  -\,   \nabla_{\mu}\phi^\dagger \nabla^{\mu}\phi\bigr )\cr
 &=\;
-eM\-P ^2R\;  +\,  {1\over2}e(\nabla_{\mu}f\nabla^{\mu}f + f^{2}
\nabla_{\mu}\theta \nabla^{\mu} \theta ) ,}\eqno (1) $$ where $\phi
={1\over\sqrt2}fe^{i\theta}$, $M\-P$ is the Planck mass\footnote{$^\dagger$}{In
the following, we explicitly retain $M\-P = (16\pi G)^{-{1\over2}}$ in our
equations, while setting $\hbar  = c= 1$.}, and $R$ is the Ricci scalar.

We shall look for wormholes using the spherically symmetric {\it ansatz}
$$
f=f(r)\qquad\qquad\theta =\theta (r)\qquad\qquad
ds^2={dr^2\over h(r)^2} + r^2d\Omega\-3^2\; , \eqno(ans)
$$
where $d\Omega\-3^2$ is the line element on the round unit three-sphere.
When we refer to any fields in the following, we will mean
only their spherically symmetric components as described in \(ans).

The field equation produced from variations of
$\theta$ has the form $\partial_{\mu}J^{\mu} = 0$ where $J^{\mu} = e
g^{\mu\nu} f^2\partial_{\nu}\theta$ is the conserved current density associated
with global variations of $\theta$ (\ie global phase rotations of $\phi$).
With
our ansatz, this equation may be integrated to yield
$$
hr^3f^2\partial_r\theta = iQ \eqno(6)
$$
where we have chosen an imaginary integration constant on the right. This
Euclidean charge is imaginary to describe tunneling between states with
real Lorentzian charge[\cite{cole}].
With this choice, the field equations determining $f$
and $h$ may be written as
$$ \eqalignno{
\partial_r(hr^3\partial_rf)&=-{Q^2\over hr^3f^3} &(7)\cr
12M\-P ^2r^4(1-h^2)&={Q^2\over f^2}-(hr^3\nabla_rf)^2\;.&(8)}
$$
Imaginary $Q$ also implies that (the lowest angular mode
of) the field $\theta$ is imaginary.  Various arguments can be advanced
to explain this use of an imaginary charge and field,
but the clearest involves Routhians[\cite{cliff, gold}].  In this formalism,
the cyclic variable $\theta$ is eliminated from the
path integral in favour of its conjugate momentum.   Consider the
spherically symmetric sector of the path integral for the scalar field theory
(without gravity),
$$
\langle F\vert e^{-2\pi^2\!\!\int\!dr\,{\s H}\-{sphere} }\vert I\rangle \sim
          \int \!{\s D}f\,{\s D}\,\theta\,{\s D}\,\pi\,{\s D}Q\>
          e^{-2\pi^2\!\int\!dr\,( {\cal H}{\scc  (\pi ,Q, f)}
          -i\pi\partial_rf-iQ\partial_r\theta ) }\;\; ,     \eqno(2)
$$
where $\pi$ and $Q$ are the momentum densities conjugate to $f$ and
$\theta $, respectively, and ${\cal H}$ is the Hamiltonian density.  Note that
${\s H}$ is independent of $\theta$.
We have Wick rotated $t \rightarrow -ir$. Implicitly the path integral above
(and in the following) includes wave functionals weighting the boundary values
as
is appropriate for the initial and final states (which we may assume are
standard
N-particle states).  The momentum sector of the path integral is a Gaussian,
since ${\cal H}$ is quadratic in the momenta.  Performing these momentum
integrals leaves the usual path integral involving only the fields
and the Lagrangian density[\cite{Feyn}].  In the present case, it is a simple
matter to integrate the cyclic field $\theta$ rather than $Q$ to give
$$
\langle F\vert e^{-2\pi^2\!\!\int\!dr\,{\s H}\-{sphere}}\vert I\rangle \sim
          \int\!{\cal D}f\,{\cal D}Q\>\delta (\partial_r Q)\>
          e^{i2\pi^2Q(\theta\-F - \theta\-I)}\>  e^{-2\pi^2\!\int\!dr\,{\s R}}
           \;\;,
              \eqno(routh0)
$$
where
$$
{\s R} =  {1\over2}r^3(\partial_rf)^2 + {1\over2}{Q^2\over r^3f^2}\;\;.
                                                                \eqno(routh1)
$$
With this approach, we acquire a phase factor at each of the boundaries
and a delta function forcing $Q$ to be constant for all $r$.  The remaining
path
integral over $f$ is weighted by the Routhian \(routh1).  The Euler-Lagrange
equation for stationary points of this new functional is
$$
\partial_r (r^3\partial_r f) = - {Q\over r^3f^3}\;,\eqno(ff)
$$
which is the analogue of \(7) with a fixed, flat metric (\ie $h=1$).

We will assume that the above approach can be extended to include gravity by
the following simple procedure:  add the Einstein action (including the
necessary surface term[\cite{surf}]) to the Routhian; covariantize the scalar
field theory; and insert our spherically symmetric mini-superspace ansatz \(2)
for the metric.  The truncation of the gravity sector in the final step still
allows us to derive the field equations, and to evaluate the classical action
for our wormhole configuration.
Equations \(routh0) and \(routh1) are then replaced by
$$
\langle F\vert e^{- 2\pi^2\!\int\!dr\,{\s H}\-{sphere}}\vert I\rangle \sim
          \int {\s D}h\,{\s D}f\>
          e^{i2\pi^2Q\,(\theta\-F - \theta\-I)}\>e^{-2\pi^2\!\int\!dr\,{\s R}}
              \eqno(new1)
$$
where
$$
{\s R} = 6M\-P^2\Bigl[ \partial_r(r^2h)- r(h + {1\over h})\Bigr] +
{1\over2}hr^3
       (\partial_rf)^2 + {1\over2} {Q^2\over hr^3f^2}\eqno(new2)
$$
and $Q$ is now a fixed constant.
It is easy to see that the Euler-Lagrange equations for stationary points of
\(new2) are just \(7) and \(8), as before.  The usual Lagrangian formalism
requires one to consider {\it a priori} imaginary $\theta$ arising from Eq.
\(6).  In contrast, in the Routhian approach, $\theta$ is eliminated by
integration over
the real axis.  Since an equivalent saddlepoint approximation may be derived
by either method, we can choose whichever technique we like.  The Routhian
formalism, although more rigorous, is more cumbersome.

Now we find wormhole solutions of Eqs. \(7) and \(8).
Differentiating \(8) and applying \(7) yields $\partial_r\Bigl(r^4(1-h^2)
\Bigr )=0$, which implies
$$
h^2=1-{L^4\over r^4}\;\;.\eqno(10)
$$
This metric is exactly the same as that found by Giddings and
Strominger[\cite{5}] with a different matter field.
The apparent singularity at $r=L$ is merely a co-ordinate
singularity[\cite{myers}].  The complete geometry is covered by two identical
co-ordinate
patches with $r\-\pm$ both ranging from $L$ to $\infty$.   The wormhole then
consists of two asymptotically flat regions where $r\-\pm\to\infty$,
connected at $r\-\pm = L$ by a throat with radius $L$.

Substituting \(10) into \(8) yields
$$
f\partial_rf=
     \pm{\sqrt{Q^2-12M\-P ^2L^4f^2}\over r^3\sqrt{1-(L/r)^4}}
\eqno(f6)
$$
which may be integrated to give
$$
\sqrt{Q^2-12M\-P^2L^4f^2}=\pm 6M\-P ^2L^2\left[\arccos \Bigl({L^2\over
r^2}\Bigr) +
C\right]\;,    \eqno(ff6)
$$
where $C$ is a (real) integration constant.  Defining
$$
x\equiv 6M\-P ^2L^2\left[\arccos \Bigl({L^2\over r^2}\Bigr)+
C\right]\eqno(xdef)
$$
we have
$$
f={\sqrt{Q^2-x^2}\over2\sqrt3M\-P L^2}\;\;.                     \eqno(11)
$$
In Eq. \(xdef) the positive (negative) branch of the {\sl arccos} is used on
the
$r\-+$ ($r\--$) co-ordinate patch.
Thus asymptotically at large radius,
$$x = 6M\-P^2L^2\left[C\pm{\pi\over2}\mp{L^2\over r\-\pm^2}
+{\s O}\Bigl({L^4\over r\-\pm^4}\Bigr)\right]\;,
$$
while the throat corresponds to $x=6M\-P^2L^2C$.
Although the wormhole geometry is symmetric on either side of the wormhole, in
general the scalar field $f$ is not when $C$ is non-vanishing.

We will connect the wormhole to a background field configuration using a
cut-and-paste procedure[\cite{cole}].  We cut the wormhole off at some {\it
fixed} scale $r\-0$ in both asymptotic regions.  Then we cut two three-spheres
of radius $r\-0$ out of the background and replace them with the ends of the
wormhole, taking care to match the boundary values of $f$ to the background
values $f\-\pm$ at $r\-\pm =r\-0$.\footnote{$^\dagger$}{Note that our procedure
differs from that implemented in [\cite{cole}].   There, the cut-off on either
side of the wormhole depends on the background field $f\-\pm$, and $C$ is fixed
to be zero.}   The scale $r\-0$ serves as the infrared cut-off for the wormhole
field configuration, which is necessary to avoid encountering divergences in
evaluating the action for the full wormhole[\cite{cole,ab}].  We may assume
that $r\-0$ is the ultraviolet cutoff for the effective low
energy theory. The wormhole ends thus appear as the insertions of local
operators since their internal structure
is beyond the limit of experimental resolution.
Further, we assume that $L$, the size of the wormhole, is near (but
larger than) the Planck scale\footnote{$^\ddagger$}{This assumption is, of
course, at the centre of a controversy[\cite{big}], to which we have nothing to
add.}, so
that $L^2/r\-0^2$ is an extremely small ratio.  This greatly simplifies
the following calculations.

The integration constants $L$ and $C$ are fixed by matching $f$ at $r\-\pm =
r\-0$ to the background values $f\-\pm$.  Eq. \(ff6) gives
$$ \eqalign{
12M\-P^2f\-\pm^2&=\left. {Q^2\over L^4}-36M\-P^4
        \left[C+\arccos \Bigl({L^2\over
r\-\pm^2}\Bigr)\right]^2\right\vert\-{r\-\pm=r\-0}\cr
      &\simeq {Q^2\over L^4}-36M\-P^4\bigl(C\pm {\pi \over 2} \bigr)^2 ,}
						    \eqno(12)
$$
where terms of order ${L^2\over r\-0^2}$ are dropped in the second line. We
therefore have
$$
C = {f\--^2-f\-+^2\over6\pi M\-P ^2} \eqno(13)
$$
and
$$
{Q^2\over L^4}=  M\-P^4\left[
               9\pi^2 +6\,{f\-+^2+f\--^2\over M\-P^2} +
               {(f\--^2-f\-+^2)^2\over\pi^2M\-P^4}\right]\;.  \eqno(14)
$$
Note that \(14) will only be consistent given the assumption that
$L>M\-P^{-1}$ for large $Q^2$.  A final comment on matching boundary conditions
is that in the background region beyond $r\-\pm=r\-0$ we employ the standard
Lagrangian formalism, and therefore immediately outside the cut-off surface we
must
enforce \(6), so that the background $\theta$ field is complex.

Now the integral of the Routhian \(new2) can be calculated for the wormhole
solution.  One finds
$$
2\pi^2\!\int\!dr\, {\s R} = \pi^2 Q \log\left|
            {Q+6M\-P^2L^2(C+{\pi\over2})\over Q-6M\-P^2L^2(C+{\pi\over2})}
           \>{Q+6M\-P^2L^2(C-{\pi\over2})\over Q-6M\-P^2L^2(C-{\pi\over2})}
                     \right| \eqno(15)
$$
where terms of order $(L/r\-0)^2$ have again been neglected.  We have
assumed $Q$ to be positive.  (With $Q<0$, the two ends of the wormhole would be
switched.)  To evaluate this expression,  we use \(13) and \(14), and
work  perturbatively  in $f\-\pm^2 /M\-P^2$.  The  final result is
$$
2\pi^2\!\int\!dr\,{\s R} \simeq
     -2\pi^2 Q\ln\left({f\-+\over \sqrt3 \pi M\-P}{f\--\over \sqrt3 \pi M\-P}
\right)
        + 2\pi^2 Q\left({f\-+^2\over 3 \pi^2M\-P^2}
        + {f\--^2\over 3 \pi^2M\-P^2}\right)\;  ,              \eqno(16)
$$
where terms of order $(f\-\pm /M\-P)^4$ have been ignored.  Thus to
tree-level order, the contribution of a single wormhole in the path integral
including the phase factors appearing in \(new1) becomes
$$
e^{i2\pi^2Q(\theta\-+-\theta\--)}e^{-2\pi^2\!\int\!dr\,{\s R}}\simeq
        A^q\phi\-+^q\phi\--^{\dagger q}\,(1-qA\phi\-+\phi\-+^\dagger)
(1-qA\phi\--\phi\--^\dagger)
 \eqno(17)
$$
where $\phi\-\pm = {1\over\sqrt2} f\-\pm e^{i\theta\-\pm}$ are  the background
scalar field values at $r\-\pm=r\-0$.  Since (as we noted above) the background
field $\theta$ is imaginary, $\phi^\dagger\equiv{1\over\sqrt2}fe^{-i\theta}$
is not the Hermitian conjugate of $\phi$
but an independent real field.  Upon analytic continuation back to Minkowski
space, however, $\phi^\dagger$ again denotes the usual Hermitian conjugate.
Also
$q=2\pi^2Q$ is the scalar charge quantized to take
values $q=1,2,3, ...$, and $A\equiv {2\over3\pi^2M\-P^2}$.

If we assume that the three-spheres $r\-\pm =r\-0$ can be taken as the
``effective points" $x\-\pm ^{\mu}$ in the background space-time, then the
effective path integral including a single wormhole is
$$
\int {\cal D}f{\cal D}\theta e^{-\int d^4x {\cal L}\-0}\;
A^q \phi^q \left.(1-Aq\phi^\dagger\phi)\right\vert\-{x=x\-+}\; \phi^{\dagger q}
\left.(1-Aq\phi^\dagger\phi)\right\vert\-{x=x\--}\;,
\eqno(back)
$$
where we have suppressed the gravity sector in this expression.
Translations of $x\-\pm^{\mu}$ are zero modes of this system, and so will be
integrated over upon evaluating quadratic fluctuations in the saddlepoint
approximation[\cite{1}].  Introducing an unknown normalization constant
$B\-q^2$, which
contains the 1-loop determinant for a wormhole of charge $q$, \(back)
becomes
$$ \int {\cal D}f{\cal D}\theta e^{-\int d^4x {\cal L}\-0}B\-q^2\int d^4x\-+ \,
A^{q/2}\phi^q(1-Aq\phi^\dagger\phi) \>\int d^4x\--  \,
A^{q/2}\phi^{\dagger q}(1-Aq\phi^\dagger\phi)\>.\eqno(better)
$$
One can show that $B_q\propto M\-P^4$\cit{cole}.
Further arguments can be made to the effect that accounting for many-wormhole
configurations within the dilute gas approximation leads to a modification of
the effective low energy action by terms of the form[\cite{alpha}]
$$
B\-qA^{q\over2}\int d^4x (\alpha\-q^\dagger\phi^q +
\alpha\-q\phi^{\dagger q})\,(1-Aq\phi^\dagger\phi)  \eqno(19)
$$
where $\alpha\-q$ and $\alpha\-q^\dagger$ might be thought of as creation and
annihilation operators for baby universes carrying global charge
$q$[\cite{andy}].

Our results are essentially the same as those found in Refs. [\cite{cole, ab}],
although our derivation differs.  Implicit in our suppression of the gravity
sector in \(back) and \(better) is the limit $M\-P \to \infty$, or rather that
the energy scales of interest are much lower than $M\-P$.  Thus the
wormhole-induced interactions are highly suppressed by the factors of $A\propto
{1\over M\-P^2}$ (for large values of q, and
excluding the possibility of drastic effects due to the
$\alpha$ parameter dynamics).  They remain as significant operators since they
break the global phase rotation symmetry $\phi\to e^{i\delta}\phi$, which would
be conserved in
all interactions induced by conventional perturbative
processes\footnote{$^\dagger$}{Ignoring gravity in the present case eliminates
all such {\it perturbative} processes, since in \(1) we only consider a free
massless scalar
field.  The above statement would be more meaningful for the case of two
interacting scalars, with the Lagrangian density $${\s L} = e
(|\nabla\phi\-1|^2 + |\nabla\phi\-2|^2 + \lambda |\phi\-1|^2|\phi\-2|^2 )\;,$$
for which the above analysis would proceed unchanged for both $\phi\-1$ and
$\phi\-2$, separately.  Of course, there may also be contributions from new
wormholes carrying charge for both $\phi\-1$ and $\phi\-2$.}.  We have included
the next-to-leading order wormhole operators, $\phi^{q+1}\phi^\dagger$ and
$\phi^{\dagger q+1}\phi$.  Since these interactions have a higher mass
dimension, they
are suppressed by an extra factor of $A$.  Therefore they will play an
insignificant role in the present bosonic theory, but we will find that they
are important for the supersymmetric case considered in the next section.

Recall that our evaluation of \(16) included a perturbative expansion in
$f\-\pm^2/ M\-P^2$, but for a strictly massless scalar, it would be difficult
to argue that these parameters should be small in the low energy theory.
Following References \cit{cole,ab}, one may consider our discussion to
apply to a scalar field with a small mass. With $m\, r\-0\ll 1$,
the mass can be neglected in the wormhole region, but
$f\-\pm^2/ M\-P^2<(M\-P m r\-0^2)^{-2}$ remains small in the low energy regime.
Alternatively, one can think of this expansion as a formal
device,
which is useful since it develops an expansion of wormhole-induced
operators of ascending mass dimension, and hence of decreasing significance in
the low energy theory.
A final comment is that to the order of this expansion
that we calculated, the wormhole contribution \(17) factorized into separate
operators at $x\-+$ and $x\--$.  This fact is important
in separating the effect of
the full wormhole into two local operators in \(back),
but there is no principle which
guarantees such a result.  In other cases, this factorization has been
found to fail[\cite{grin}], and indeed it fails in the present
case at the next order beyond those we have displayed.

\head{3.  The Superwormhole}

We now wish to study wormholes in a supersymmetric theory. Explicitly,
we will consider a Wess-Zumino multiplet coupled to $N=1$ supergravity.
The essential new aspect in this case is the application of saddlepoint
approximations to a theory containing fermions as well as bosons.
In this case, the fermionic zero modes will produce anticommuting
collective co-ordinates.
The specifics of the supergravity theory are irrelevant to most of our results
in this section. Therefore in the interests of clarity, we
will leave the details concerning the fermionic zero modes of the
superwormhole to Appendix A. All that we require is to note the theory
possesses the following properties:

1) If all fermion fields are set equal to zero, we recover the Lagrangian of
section 2. Therefore the wormhole solution of that section is also a
saddlepoint of the present theory.

2) In the limit $M\-P \to \infty$ we obtain the massless Wess-Zumino
model, with vanishing super-potential, in flat space.  This fixes the
form of the supersymmetry transformations in the following.

3) The theory is $N=1$ supersymmetric in Euclidean four-space. Thus when
fermionic zero modes arise in the fluctuation determinant of the saddlepoint,
the corresponding collective co-ordinate is a Grassman four-spinor $\eta$.

In Appendix A, we find fermionic zero modes in the wormhole background,
which arise because of the invariance of the action under
supersymmetry transformations. There are four independent modes associated
with each end of the wormhole. Each mode is equivalent to a global Wess-Zumino
supersymmetry transformation in one asymptotic region, and vanishes in
the opposite asymptotic region.
These zero modes (as well as the bosonic ones) are separated, and the
saddlepoint expansion is performed only on the remaining modes of the path
integral[\cite{raja}].

Let $\Theta(\veps)$ be the charge generating the Wess-Zumino supersymmetry
transformation of property 3 parameterized by
an arbitrary Grassman parameter $\veps$.
The fermion zero modes $\eta$ may be separated
by constructing the wormhole path integral with identity
operators $e^{-i\Theta(\eta)}\,e^{i\Theta(\eta)}$
inserted between the time-slices. To leading
order in the saddlepoint expansion, we obtain the
the bilocal effective interaction
$$
{\hat B}_q^2\int d^4x\-+d^4x\--d^4\eta\-+d^4\eta\--\ e^{-i\Theta(\eta\-+)}
O\-{w}(x\-+)\,O\-{w}^\dagger(x\--)\ e^{i\Theta(\eta\--)}
\ ,\eqno(effint)
$$
where $O\-{w}\equiv A^q\phi^q \,(1-Aq\phi^\dagger\phi)$ is the bosonic
operator found in section 2, and $\eta\-+(\eta\--)$ are the fermion
zero modes, which are constant in the asymptotic region with
$r\-+(r\--)$ but vanish for $r\--(r\-+)\rightarrow\infty$.
The result of the fluctuation determinant is contained in ${\hat B}_q^2$.
In this case, we may assume as a result of property 3 that
the fermionic and bosonic
determinant terms with all zero modes extracted cancel each
other, but there will be factors arising from the normalization of the
zero modes. In particular, $\hB_q$ will have the dimensions of mass
squared.

Above, we have taken the Euclidean time to
flow radially through the wormhole, increasing in the direction of growing
$x$ or $r\-+$. We relate our result \(effint) to the choice
where time flows in a fixed direction across the background regions
(see Figure 1) as follows:
The wormhole path integral with radial time produces an evolution
operator from the surface I (at $r\-- = r\-0$) to F (at $r\-+ = r\-0$).
This operator is inserted between the
surfaces $i\-\pm$ and $f\-\pm$ in the background  spaces.
The integrals of the supercurrent over I and F yielding the charges
$\Theta(\eta\-\pm)$  are then split into two integrals over these
background surfaces.
After baby universe $\alpha$ parameters are used to localize the effective
interaction as in section 2, the above procedure yields an effective
operator for the superwormholes of the form
$$
\hB_qA^{q/2}\int\! d^4x\,d^4\eta\ e^{-i\Theta(\eta)}(\alpha\-q^\dagger\phi^q +
\alpha\-q\phi^{\dagger q})\,(1-Aq\phi^\dagger\phi)\,
e^{i\Theta(\eta)} \ .\eqno(yes!)
$$
This is clearly a superspace vertex[\cite{6}], so that as expected
the superwormhole terms manifestly preserve the supersymmetry of the low energy
background theory. From here onwards, the superspace formalism provides the
most elegant framework for describing the low energy limit of superwormholes.

The massless Wess-Zumino model without a superpotential has the simple
Euclidean action
$$
{\s S}\-{WZ}\ = \int\! d^4x\,[
    \delta^{\mu\nu}\partial_{\mu}\phi^\dagger\partial_{\nu}\phi
         + {1\over2} \bar\chi\slap\chi
         - F^\dagger F]\;, \eqno(wzlag)
$$
where $\delta^{\mu\nu}$ is the flat Euclidean metric, and
$\slap=\gamma^{\mu}\partial_{\mu}$.
The matter fields have some unfamiliar characteristics
due to the analytic continuation required to implement supersymmetry in
Euclidean four-space. (Appendix A discusses this point at length.)
The essential point is that because there are no Majorana spinors
in Euclidean four-space, the adjoint spinors are defined as
$\bar\chi\equiv\chi^T C$. As a result, one must again think of
$\phi$ and $\phi^\dagger$ (and $F$ and $F^\dagger$, as well) as independent
fields.
${\s S}\-{WZ}$ is invariant under the following global supersymmetry:
$$\eqalign{
\delta \phi & = \sqrt2 \bar\veps P\-L \chi\qquad\qquad\ \ \qquad\qquad
\delta \phi^\dagger =\sqrt2 \bar\veps P\-R \chi\cr
\delta \chi & = \sqrt2 P\-L (\slap\phi + F)\veps
              + \sqrt2 P\-R(\slap\phi^\dagger+F^\dagger)\veps\cr
\delta F & = \sqrt2 \bar\veps\slap P\-L\chi\qquad\qquad\qquad\qquad
\delta F^\dagger =\sqrt2 \bar\veps\slap P\-R\chi\;,}\eqno(fsusy)
$$
where $P\-{L/R} \equiv{1\over2}(1\pm\gamma\-5)$.

The auxillary fields, $F$ and $F^\dagger$, are decoupled in \(wzlag), and
can be trivially integrated out.
Their virtue is that they allow the supersymmetry transformations \(fsusy) to
close off-shell. Therefore when the action is altered by the addition of
our wormhole terms, the supersymmetry transformations remain unchanged if we
include $F$.
The supersymmetries of the free and wormhole-modified actions would
differ if expressed in terms of physical fields only.
The scalar superfield $\Phi(x^{\mu},\veps)$ may be defined as the image
of $\phi(x^{\mu})$ under a finite supersymmetry transformation:
$$\eqalign{
\Phi&(x,\veps) \equiv  e^{-i\Theta(\veps)}\phi(x)\,e^{i\Theta(\veps)}\cr
 &=\phi +\sqrt2\bar \veps P\-L \chi +(\bar \veps P\-L \veps )F +
(\bar \veps P\-L\gamma^{\mu}\veps )\partial_{\mu}\phi
+{1\over\sqrt2}(\bar\veps P\-L \veps )\bar \veps \sla
\partial P\-L \chi +{1\over8}(\bar \veps \veps )^2 \part^\mu\part_\mu\phi\; .}
\eqno(spfd)
$$
Similarly $\Phi^\dagger\equiv e^{-i\Theta(\veps)}\phi^\dagger
e^{i\Theta(\veps)}$.
Since the adjoint spinor $\bar\veps$ is a Majorana conjugate, the conjugate
superfield $\Phi^\dagger$ is once again not the true complex conjugate
of $\Phi$. Its definition though does replace every $P\-L$ in \(spfd)
with $P\-R$, and every $\phi$ with $\phi^\dagger$.

The Wess-Zumino action \(wzlag) can be written as a superspace
integral\footnote{$^*$}{One uses the standard convention for Grassman
integration: $\int\!d\veps\,[a\veps +b] = a$ for
a Grassman $\veps$, and ordinary numbers $a$ and $b$.}
$$
{\s S}\-{WZ} = -{1\over4}\!\int\!d^4x\,d^4\eta\ \Phi^\dagger(x,\eta)
\Phi(x,\eta)\;\;.\eqno(s0)
$$
By applying \(spfd) to \(yes!), we see that the
superwormholes contribute extra vertices to this superspace action:
$$
\hB_qA^{q/2}\!\int\!d^4x\,d^4\eta\,(1-Aq\Phi^\dagger\Phi)
(\alpha\-q^\dagger\Phi^q\,+\,\alpha\-q\Phi^{\dagger q})\;\;
.\eqno(spwmh)
$$
In the end then, our result from section 2 has been supersymmetrized
in the most obvious way.

\head{4. Conclusion}

Combining \(s0) and \(spwmh), we see that the effect of the superwormholes
is to add extra terms to the Kahler potential of the Wess-Zumino
multiplet.\footnote{$^\ddagger$}{In this discussion, it will be
assumed that we have rotated back to Lorentzian signature, and so the
fields have no unusual characteristics.}
The $\Phi^q$ and ${\Phi}^{\dagger q}$ terms can be removed by a Kahler gauge
transformation\cit{twilight}. Alternatively,
integrating over $\eta$ in \(spwmh), one finds explicitly that these terms
contribute only total derivatives. Thus as expected\cit{k},
supersymmetry suppresses the wormhole-induced scalar
self-couplings.  This leaves
$$  V(\Phi,\Phi^\dagger) =
 -{1\over4}\Phi^\dagger\Phi\; +\;\hB_q \sum_{q>0}qA^{{q\over2}+1}
(\alpha^\dagger\-q\Phi^{q+1}\Phi^\dagger + \alpha\-q\Phi^{\dagger(q+1)}\Phi)
\eqno(seff)
$$
as the tree-level Kahler potential incorporating the effects of
superwormholes. Explicitly in terms of the physical fields,
the Lagrangian density is
$$\eqalign{
{\cal L} =& -4\,\parb\part V\,(\nabla_{\!\mu}\phi\nabla^\mu\phi^\dagger+
{1\over2}\bar\chi\sla\nabla\chi)\cr
&+\,\bar\chi\gamma\-5\gamma^{\mu}\chi\,(\parb\part^2\!V\,\nabla_{\!\mu}\phi-
\parb^2\!\part V\,\nabla_{\!\mu}\phi^\dagger)+{1\over2}\!\left(
\parb^2\!\part^2\!V
-{|\parb\part^2\!V|^2\over\parb\part V}\right)(\bar\chi\chi)^2\cr}
\eqno(stuffl)$$
where $V=V(\phi,\phi^\dagger)$, $\part\equiv{\delta\ \over\delta\phi}$, and
$\parb\equiv{\delta\ \over\delta\phi^\dagger}$.
For a given charge $q$, the leading superwormhole
induced terms are suppressed by an extra factor of $M\-P^{-4}$, as compared
to the purely bosonic theory of Section 2.

Even these latter terms are evanescent, in fact.
We can absorb the superwormhole vertices to linear order in the
$\alpha$ parameters, with a holomorphic field redefinition
$$ \tilde\Phi = \Phi -
4\sum_{q>0}qA^{{q\over2}+1}\alpha^\dagger\-q\Phi^{q+1} \;. \eqno(redone)$$
Note that this field redefinition is $\alpha$-{\it dependent}.
We now recover the free Wess-Zumino model,
up to terms quadratic in baby universe parameters
$$
\widetilde{V}(\tilde\Phi,\tilde\Phi^\dagger) =
-{1\over4}\tilde\Phi^\dagger\tilde\Phi - 4\sum_{q>0}\sum_{q'>0}
qq'A^{{q+q'\over2}+2}\alpha^\dagger\-q\alpha\-{q'}\tilde\Phi^{q+1}\tilde
\Phi^{\dagger q'+1}+{\s O}(\alpha^3)\;\;.\eqno(supp)
$$
Both Kahler potentials, \(seff) and \(supp), will produce an equivalent
physical theories. Therefore,
because the leading symmetry-breaking interactions in \(stuffl)
can be eliminated by field
redefinitions, these terms  will not directly affect physical
scattering processes, which might display violations of charge
conservation. Terms quadratic and higher order in the $\alpha$ parameters,
are already present in \(stuffl) in the last term, which was produced
by integrating out the auxillary fields, $F$ and $F^\dagger$.
A typical process with a charge violation of $\pm q$ units might
then be mediated by interactions of the form
$$ (\alpha\-1\alpha^\dagger\-{q+1}\,\phi^q+
\alpha^\dagger\-{1}\alpha\-{q+1}\,\phi^{\dagger q})\,(\bar\chi\chi)^2\ .
\eqno(forrm)$$
These superwormhole induced interactions are now
suppressed by an extra factor of $M\-P^{-6}$, as compared
to the purely bosonic theory.

It may seem curious that symmetry-breaking first occurs at order
$\alpha^2$. It appears then that the observable effects are only occurring in
multi-wormhole processes. This result is analogous to certain
instanton effects
found in Reference \cit{ian}. There, in a particular (2+1)-dimensional
gauge theory, the photon and photino are found to acquire masses
only through
contact terms arising in multi-instanton processes. One should not think
that a charge-$q$ superwormhole (which has four fermionic zero modes) must
induce interactions of the form \(forrm) directly. This is because the
fermionic zero modes are not strictly zero modes of the fermion fields,
since they are extracted using finite (nonlinear) supersymmetry
transformations, which involve the bosonic fields as well.

Ultimately though, we expect that interactions like \(forrm)
should arise in the single superwormhole sector. Above, we have ignored the
possibility that the Kahler potential might contain higher dimension
interactions such as
$$  \delta V(\Phi,\Phi^\dagger) =
 M\-P^2 \sum_{n>1}C_n \left({\Phi\,\Phi^\dagger\over M\-P^2}\right)^n
\ ,\eqno(seff1)
$$
where $C_n$ are dimensionless constants.
Such terms are present in the full supergravity theory, but would also
arise in the usual renormalization of the Kahler potential, when modes
at wavelengths shorter than $r\-0$ are integrated out. So far, such terms
have been neglected on the basis of the arguments presented in section 2:
they obey the phase rotation symmetry,
and their effects should be
suppressed at low energies because they yield higher
dimension operators in \(stuffl).
In fact though, they play a
significant role in the charge violating processes in the supersymmetric
theory. Applying the field
redefinition \(redone) to $V$+$\delta V$ leaves symmetry breaking
terms linear in the $\alpha$ parameters.
Now a typical process with a charge violation of $\pm q$ units could
be mediated by interactions of the form
$$ C\-2\,(\alpha^\dagger\-{q}\,\phi^q+
\alpha\-{q}\,\phi^{\dagger q})\,(\bar\chi\chi)^2\ .
\eqno(forrm1)$$
Since these interactions
are linear in the $\alpha$ parameters, they produce symmetry breaking
processes in a single superwormhole background. Note that they are
still second order in a combined perturbation expansion in terms of
the $\alpha$'s and $C$'s.
Of course, the new terms have the same dimension as those given in \(forrm),
and so supersymmetry supresses the observable wormhole effects in
any event.

Note that before wormhole effects are taken into account, the Wess-Zumino
theory has two independent global U(1) symmetries:
$$
\phi \to e^{i\delta}\phi \qquad\qquad{\rm and}\qquad\qquad
\chi\to e^{i\gamma_{5}\lambda}\chi\,. \eqno(uuuu)
$$
Examining the form of the interactions in \(stuffl), we see that the
wormhole induced terms only break the phase rotation symmetry of
the scalar field. The fermion's chiral rotation symmetry remains
unbroken. One might expect that the latter symmetry must also
be broken, since in the full supergravity theory, the global $U(1)$
symmetry requires $\delta=\lambda$. (This comes about from couplings of the
matter fields with the gravitino.) In fact though, the chiral rotations
are independent because of the $R$-symmetry of the supergravity theory:
$\chi\to e^{i\gamma_{5}\lambda}\chi,$
$\psi_\mu\to e^{-i\gamma_{5}\lambda}\psi_\mu,$
$\epsilon\to e^{-i\gamma_{5}\lambda}\epsilon.$
This chiral symmetry may be broken by
new wormholes in which the U(1) charge is carried by
the fermions\cit{kim}.

Finally we observe that our superwormholes do not violate the
nonrenormalization theorem for chiral superfields\cit{twilight}.
This failure to induce a superpotential, means that these wormholes
cannot produce spontaneous supersymmetry breaking.
Essentially this occurs because there are four
independent fermionic zero modes for each end of the wormhole,
which leads to the $\int d^4\!\eta$ in the effective local operators.
A superpotential requires an F term, which would only
contain a chiral Grassman integration, $\int d^2\!\eta$.
Such a result was found for a particular wormhole\cit{sstev} in
Reference \cit{10}. In this case, although there are four fermionic
zero modes, two are not normalizable and so do not contribute.
At present, this effect appears to depend on the detailed dynamics of their
theory. It would be interesting if any general statements could be
made as to which theories or wormholes would yield such a result.

\bigskip
\bigskip

The authors would like to thank Cliff Burgess and Andy Strominger for useful
conversations.  This research was supported by NSERC of Canada, and by Fonds
FCAR du Quebec.  R.C.M. would like to thank the Institute for Theoretical
Physics at UCSB and the Aspen Center for Physics for their hospitality at
various stages of this work.  At UCSB, this research was also supported in part
by NSF Grant PHY 89-04035.

\bigskip
\bigskip
\bigskip
\bigskip

\head{Appendix A.}

In this appendix, we explicitly calculate the fermion zero modes of
the wormholes in the supergravity model.
Furthermore, we demonstrate the three
properties on which the results of section 3 depend.

To study a superwormhole, we must extend the field theory of section 2 to a
Wess-Zumino multiplet coupled to
N=1 supergravity.  The Lorentzian Lagrangian for such a system has been
determined[\cite{bagger, cremmer}]; the simplest case (\ie that with canonical
kinetic terms) may be written as

$$\eqalign{ {\cal L}\-{L} =\; -M\-P^2eR\;
   -\,{1\over2}e(\nabla_{\!\mu}f\nabla^{\mu}f\;
   +\;f^2\nabla_{\!\mu}\theta&\nabla^{\mu}\theta)\cr
-{1\over2}(i\bar\psi_{\mu}\gamma\-5\gamma_{\nu}\nabla_{\rho}\psi_{\sigma}
                                              & \eps^{\mu\nu\rho\sigma}\;
   +\;e\bar\chi\sla\nabla\chi)\cr
+\,{1\over 2\sqrt2 M\-P}e\bar\chi
   \gamma^{\mu}\gamma^{\nu}\nabla_{\nu}(fe^{-i\gamma\-5\theta})\psi_{\mu}
-\,{1\over16M\-P^2}f^2&\nabla_{\!\mu}\theta
   (\bar\psi_{\nu}\gamma_{\rho}\psi_{\sigma}\eps^{\mu\nu\rho\sigma}
   +ie\bar\chi\gamma\-5\gamma^{\mu}\chi)\cr
   +\,{1\over 128M\-P^2}
   \bar\chi\gamma\-5\gamma_{\sigma}\chi\,(4i\bar\psi_
   {\mu}\gamma_{\nu}\psi_{\rho}\eps^{\mu\nu\rho\sigma}
   -4&
      e\bar\psi_{\mu}\gamma\-5\gamma^{\sigma}\psi^{\mu}
   -e\bar\chi\gamma\-5\gamma^{\sigma}\chi )\;.
}\eqno(lag0)
$$
The matter fields are: $f$, a real scalar field; $\theta$, a real periodic
pseudo-scalar; and $\chi$, a Majorana spinor. In the gravity sector, one has
the gravitino, $\psi_{\mu}$, which is a Majorana spinor-vector; and
the vierbein, $e^a{}_{\mu}$.\footnote{$^\ddagger$}{Greek and Roman letters %
indicate tensor and Lorentz indices, respectively. The latter are raised and %
lowered with the trace +2 Minkowski metric.}
The spin connections in all the covariant derivatives are the usual connections
compatible with the vierbein, plus torsion terms involving the
gravitino[\cite{West}].  Using these spin connections, the Ricci scalar is
defined from the Riemann tensor: $R \equiv e^{\mu}_a
e^{\nu}_b R_{\mu\nu}{}^{ba}$.  Finally, $e\equiv|{\rm det}\, e^a{}_{\mu}|$,
and  $\epsilon^{\mu\nu\rho\sigma}$ is the antisymmetric tensor {\it density},
defined so that $\epsilon^{0123} = 1$.

One may choose a real representation for $\gamma^\mu$, in which the
Majorana spinor fields are real, and
$\gamma\-5$ is imaginary.  This convention will be useful below in making clear
our method of Wick rotating \(lag0).

	This Lagrangian is invariant (up to a total derivative) under the
following local supersymmetry transformations\cit{bagger, cremmer}:
$$
\eqalign{
\delta f\>\>\;&=\;\bar\epsilon e^{-i\gamma\-5\theta}\chi ;\qquad
\delta \theta\>\>\;=\;-{i\over f}\bar\epsilon\gamma\-5 e^{-i\gamma\-5\theta}
                        \chi ;\cr
\delta\chi\>\>\;&=\;e^{-i\gamma\-5\theta}(\nabla_{\mu}f +
                if\gamma\-5\nabla_{\mu}\theta )
	\gamma^{\mu}\epsilon\;+\;{f\over8 M\-P^2}
(\bar\chi e^{-i\gamma\-5\theta}\gamma\-5\epsilon )\gamma\-5\chi\cr
&\qquad\qquad-\;{1\over2\sqrt2 M\-P}\bigl ((\bar\psi^{\mu}
\chi)\>+\gamma\-5(\bar\psi^{\mu}\gamma\-5\chi)\bigr )\gamma_{\mu}
\epsilon ;\cr
\delta e_{\mu}^a\;&=\;{1\over \sqrt2 M\-P}\bar\epsilon\gamma^a\psi_{\mu};\cr
\delta\psi_{\mu}\>\;&=\;2\sqrt2 M\-P
\nabla_{\mu}\epsilon\>-\>{f\over 8M\-P^2}
(\bar\chi e^{-i\gamma\-5\theta}\gamma\-5\epsilon)
\gamma\-5\psi_{\mu} \cr &\qquad\qquad +{1\over 4\sqrt2M\-P}
\bigl (\sigma_{\mu\nu}(\bar\chi\gamma\-5\gamma^{\nu}\chi)\>+\>
2if^2\nabla_{\mu}\theta\bigr )\gamma\-5\epsilon \>; }\eqno(lsusy0)
$$
to first order in the Grassman Majorana spinor field $\epsilon$.
As usual, $\sigma_{\mu\nu}\equiv {1\over2}[\gamma_{\mu},\gamma_{\nu}]$.

In order to find a wormhole, we must first Wick rotate to
Euclidean four-space.  This poses an apparent problem, because there are no
Majorana spinor representations of SO(4).
However, while we cannot find spinor representations of
SO(4) such that $\chi^{\dagger}\gamma_4 = \chi^TC$, they are not needed
--- actually we only need
$$
\bar\chi =\chi^TC\ .\eqno(conj)
$$
Therefore we {\it define} our adjoint spinors with Majorana conjugation
\(conj)\cit{majic}. In the Lorentzian theory where the fermions are
Majorana spinors,
the use of this convention instead of the usual Dirac conjugation does not
affect the theory. In the Euclidean version of the
theory, the action will only contain $\chi$ and $\psi_\mu$, making
no reference to $\chi^\dagger$ or $\psi_\mu^\dagger$. Hence the fermion path
integral can be regarded as an analytic contour integral in
the spinor field space, and still contains precisely the correct number
of degrees of freedom for a supersymmetric theory.

Wick rotation therefore yields
$$\eqalign{ {\cal L}\-{E} =\; -M\-P^2eR\;
   +\,{1\over2}e(\nabla_{\!\mu}f\nabla^{\mu}f\;
   +\;f^2\nabla_{\!\mu}\theta&\nabla^{\mu}\theta)\cr
-\,{1\over2}(\bar\psi_{\mu} \gamma\-5\gamma_{\nu}\nabla_{\!\rho}
   \psi_{\sigma}&\epsilon^{\mu\nu\rho\sigma}
    \,-\,e\bar\chi\sla \nabla \chi) \cr
-\,{1\over 2\sqrt2 M\-P}e\bar\chi
   \gamma^{\mu}\gamma^{\nu}\nabla_{\!\nu}(fe^{-i\gamma\-5\theta})\psi_{\mu}
+\,{i\over16M\-P^2}f^2&\nabla_{\!\mu}\theta
   (\bar\psi_{\nu}\gamma_{\rho}\psi_{\sigma}\epsilon^{\mu\nu\rho\sigma}
   +e\bar\chi\gamma\-5\gamma^{\mu}\chi)\cr
+\,{1\over 128M\-P^2}
   \bar\chi\gamma\-5\gamma_{\sigma}\chi\,(4\bar\psi_
   {\mu}\gamma_{\nu}\psi_{\rho}\epsilon^{\mu\nu\rho\sigma}
   +4e&\bar\psi_{\mu}\gamma\-5\gamma^{\sigma}\psi^{\mu}
   +e\bar\chi\gamma\-5\gamma^{\sigma}\chi )\;.
}\eqno (lag1)
$$
Our conventions are: $\epsilon^{\mu\nu\sigma\rho}$ is $\pm1$ or 0
in both Euclidean and Lorentzian space.  Euclidean indices range from 1 to 4
instead of 0 to 3. $v^4\equiv -iv^0\,,v_4\equiv iv_0$ where the vector $v$
has either tensor or SO(4)/Lorentz indices.
Note that $\s L_{E}$ is not real, but this is not a problem --- we only require
a Hermitian Lagrangian when we analytically continue back to real time.
Note that \(lsusy0) is indeed a symmetry of ${\cal L}\-E$ using
Majorana conjugation to define the adjoint spinors. However if the
phase of $\epsilon$ is arbitrary, then these transformations may not
preserve the reality of all the bosonic fields in the Euclidean theory.
If we choose the phase to always preserve the reality of
the vierbein, one must regard the matter fields
$\phi={1\over\sqrt2}fe^{i\theta}$ and $\phi^\dagger={1\over\sqrt2}f
e^{-i\theta}$ as independent fields rather than complex conjugates.

It is immediately clear that setting $\chi$ and $\psi_{\mu}$ equal to zero in
\(lag1) produces the bosonic Lagrangian \(1).  This establishes property
1 of section 3, which showed that the bosonic wormhole solution is also
a saddlepoint of the supergravity theory. Also,
since \(lsusy0) is still a symmetry of the Euclidean Lagrangian,
property 3 is valid ($N=1$ supersymmetry).

We can now proceed with a supersymmetric version of the Routhian formalism.
Our system is invariant under the global U(1) transformation
$$
\eqalign{\theta &\to \theta + \delta\cr
\chi\;&\to\;e^{i\gamma_{5}\delta}\chi\,.} \eqno(28)
$$
This symmetry provides the generalization
of the charge that was important in section 2.  We can simplify many of our
subsequent expressions by re-defining the fermion field so that it is invariant
under \(28). Henceforth, we use
$$
\chi = e^{i\gamma\-5\theta}\xi .\eqno(xi)
$$
The Lagrangian may be re-written as
$$
{\cal L}\-{E} \equiv\;\bar\s L\;
+\,e\beta^{\mu}\nabla_{\mu}\theta\;
+\,{1\over2}ef^2\nabla_{\mu}\theta\nabla^{\mu}\theta\; ,\eqno(30)
$$
where $\bar\s L$ is independent of $\theta$:
$$
\eqalign{ \bar{\cal L} =\; -M\-P^2eR\;  +\,{ {1\over2}}
   e\nabla_{\mu}f\nabla^{\mu}f\; +\,{ {1\over2}}&
                                            e\bar\xi\sla \nabla
   \xi \; -\, {1\over2}\bar\psi_{\mu} \gamma\-5\gamma_{\nu}\nabla
   _{\rho}\psi_{\sigma}\epsilon^{\mu\nu\rho\sigma} \cr
-\,{\nabla_{\nu}f\over 2\sqrt2 M\-P}e\bar\xi
   \gamma^{\mu}\gamma^{\nu}\psi_{\mu}\;&+\,{e\over32M\-P^2}(\bar\xi\xi)^2\cr
   +\,{1\over 32M\-P^2}\bar\xi\gamma\-5\gamma_{\sigma}\xi\,(\bar\psi_{\mu}
   \gamma_{\nu}\psi_{\rho}&\epsilon^{\mu\nu\rho\sigma}\;+\,
                                                        e\bar\psi_{\mu}
   \gamma\-5\gamma^{\sigma}\psi^{\mu})\;,}\eqno(30)
$$
and
$$
\beta^{\mu}\equiv\;
{if^2\over 16M\-P^2}
({1\over e}\bar\psi_{\nu}\gamma_{\rho}\psi_{\sigma}\epsilon^{\mu\nu\rho\sigma}
\;+\,e\bar\xi\gamma\-5\gamma^{\mu}\xi)\;
-\,{i\over2}\bar\xi\gamma\-5\gamma^{\mu}\xi
+\,{i\over2\sqrt2M\-P}\bar\xi\gamma^{\nu}\gamma^{\mu}\gamma\-5\psi_{\nu}\,.
\eqno(beta)
$$

We can now construct a Routhian form of the path integral following
\reff{cole}. The supergravity Routhian is obtained by introducing the
four-vector density $j^{\mu}$ conjugate to $\theta$, which allows us to
exchange the path integral over $(e_{\mu}^a,\psi_{\mu},\xi,f,\theta)$ for
one over $(e_{\mu}^a,\psi_{\mu},\xi,f,j^{\mu})$,
$$\eqalign{
\langle F| e^{-\int\!d^4x\,{\s H}}|I\rangle\; =\;&
\int\!{\s D}e_{\mu}^a\,{\s D}\psi_{\mu}\,{\s D}\xi\,{\s D}f\,{\s D}\theta\;\;
e^{-S}\cr
=\;&\int\!{\s D}e_{\mu}^a\,{\s D}\psi_{\mu}\,{\s D}\xi\,{\s D}f\,{\s
D}j^{\mu}\;\;
e^{-S'}\;\;.}
$$
We multiply the path integral by a normalized
Gaussian integral over $j^{\mu}$.
$$\eqalign{
e^{-S}\; &= \int{\s D}j^{\mu}
		\;e^{-\int d^4x {\s L}}\;e^{-\int d^4x {1\over 2f^2e}
		[j^{\mu}-ie(f^2\nabla^{\mu}\theta+\beta^{\mu})]^2}\cr
                           &= \int {\s D}j^{\mu}
                e^{-\int d^4x [\bar{\s L} + {1\over2ef^2}
              (j^{\mu}-ie\beta^{\mu})^2 - ij^{\mu}\nabla_{\mu}\theta]}
                }\eqno(ro0)
$$
As in section 2, the term $j^{\mu}\nabla_{\mu}\theta$ is integrated by
parts, and then the path integral over $\theta$ produces a delta
function and surface terms.
$$
e^{-S'} = \exp\left[-i\oint d^3\Omega\-3 Q(\theta\-+-\theta\--)  \;-\,
\int\! d^4x\, {\s R}\right]\times
\delta(\partial_{\mu}j^{\mu})\;  \eqno(32)
$$
where the Routhian $\cal R$ is
$$
{\cal R} =\;\bar\s L + {1\over2f^2} (j_{\mu} -
 ie\beta_{\mu})^2\;. \eqno(blap)
$$
Also $\theta\-+$ and $\theta\--$ are values of $\theta$ on boundary surfaces,
and $Q$ is the U(1) charge passing through these surfaces.
As in section 2, the initial and final slices are
the cut-off surfaces on either side of the superwormhole, at $r_\pm=r\-0$.
Also, we will employ $e^{-S'}$ inside the wormhole region, and $e^{-S}$ in
the background outside.  Summing over all values of the charge $Q$
makes the surface terms in \(32) serve as projection operators from the
$\theta$ basis to the basis of charge eigenstates\cit{cole}.

We must now consider the supersymmetry properties of our Routhian.
These are most easily determined by considering the equations of motion.
The saddle-point of the Gaussian in \(ro0) is given by
$$
j_{\mu} = ie(f^2\nabla_{\mu}\theta + \beta_{\mu})\ .\eqno(zm0)
$$
Up to the substitution \(zm0), the equations of motion (and hence the
saddle-points) of
${\s R}$ and ${\s L}$ coincide.
Having made the substitution \(xi), the only appearance
that $\theta$ makes in the supersymmetry transformations \(lsusy0)
is as $\nabla_{\!\mu}\theta$. These terms can be replaced using \(zm0),
and the supersymmetry transformation of $j_\mu$ is easily
derived as the variation of the right-hand-side of \(zm0) under
\(lsusy0).

The local supersymmetry transformations describe the
fermionic gauge degrees of freedom of the supergravity theory.
To discuss the physical degrees of freedom, one must prescribe a
particular choice of gauge-fixing. We will impose $\gamma^{\mu}\psi_{\mu}
= 0$. Given the wormhole solution, certain transformations
will still leave the gauge constraint invariant. These field variations
are the physical (fermionic) zero modes.
The zero mode equation, $\gamma^{\mu}\delta\psi_{\mu} =0$, yields
in the bosonic wormhole background,
$$
\sla\nabla\epsilon = - {Q\over 4M\-P^2 r^3}\gamma^4\gamma\-5\epsilon
\ .\eqno(zma)
$$
More convenient co-ordinates are those that map both sides of the
wormhole.  Let $x^4=x\equiv\arccos {L^2\over r^2}$, and take Euler angles
on the three-sphere as $x^i$ for
$i\in\{1,2,3\}$.  Using the vierbien and spin connection for the metric \(ans)
transformed into these co-ordinates, \(zma) becomes
$$
\gamma^4 \left( 2(\cos{x})^{3\over2}\partial_x +
                  {3\over2}(\cos{x})^{1\over2}\sin{x}\right)\epsilon
-(\cos{x})^{1\over2}\gamma^i\nabla_i^{(3)}\epsilon - {Q(\cos{x})^{3\over2}
\over8M\-P^2L^2}\gamma^4
\gamma\-5\epsilon = 0 ,\eqno(zm1)
$$
where $\nabla_i^{(3)}$ is the spinor covariant derivative on the round unit
three-sphere, and $\gamma^i$ are Dirac matrices projected onto the unit
three-sphere dreibein.  We can separate variables by choosing the {\it ansatz}
$$
\epsilon = k(x)\eta (\Omega )
$$
with $\eta$, a spinor that depends only on the angular variables,
 and the matrix $k(x)$ takes the form, $k = k\-1 + k\-5\gamma\-5$.

Asymptotically we are looking for a spinor which becomes covariantly constant,
so that the variation of the fields vanishes there.
$$
\lim_{x\to \pm{\pi\over2}} \nabla_{i}\epsilon = \lim_{x\to\pm{\pi\over2}}
(\nabla_i^{(3)}\epsilon + {1\over2}\sin{x}\gamma^4\gamma_i\epsilon) = 0
       \eqno(zm2)
$$
This implies that
$$
\gamma^i\nabla_i^{(3)}\eta = \pm {3\over2}\gamma^4\eta
                                                                    .\eqno(zm3)
$$
Notice that the only way both limits in \(zm2) can hold true is
if $\epsilon$ vanishes at one end of the wormhole, and approaches a constant
$\eta$ at the other.  The equation for $k$ is
$$
\partial_x \ln k = {Q\over 16M\-P^2L^2}\gamma\-5 - {3\over4}
  \tan{\left({x\over2}\pm{\pi\over4}\right)},
$$
which is easily solved to yield
$$
\epsilon_\pm
 = \left[\cos{\left({x\over2}\pm{\pi\over4}\right)}\right]^{3\over2}
   \exp\left[{{Q\gamma\-5(x\pm{\pi\over2})\over12M\-P^2L^2}}\right]\,\eta
\ .\eqno(zm4)
$$
As in section 2, there are distinct zero modes for each end of the wormhole.
$\epsilon_\pm$ vanish for $x\rightarrow\pm{\pi\over2}$
and become covariantly constant spinors $\eta$ as $x\rightarrow\mp{\pi\over2}$.
Being covariantly constant according to the flat space connection which
prevails
asymptotically far from the wormhole, it is $\eta$ which is analogous
to the constant parameter of rigid supersymmetry, and which
parametrizes the supersymmetric zero mode of the wormhole.

Now finally we return to property 2 of section 3.
Letting $M\-P\to\infty$ in \(lag1) (and
neglecting free gravitons and gravitinos) produces the Euclidean Lagrangian of
the free Wess-Zumino model.
Furthermore in this limit with a covariantly constant supersymmetry parameter,
\(lsusy0) reduces
to the standard supersymmetry transformations of the free Wess-Zumino
model. Since our zero modes are indeed covariantly constant asymptotically,
they correspond precisely to a global Wess-Zumino supersymmetry
transformation, and property 2 is also valid.
Since the vierbein and gravitino are invariant under these
transformations, we are justified in disregarding the supergravity sector
in the background
region, and representing the effects of wormholes as extra vertices in an
effective {\it flat} superspace action. Above we have only considered the
first order supersymmetry transformations. The finite transformations may be
obtained by iterating our first order calculation up to fourth order. It
is straightforward to show that, through all iterations, the fermionic
zero mode still corresponds to the Wess-Zumino supersymmetry to leading
order in $L/r\-0$.

\vfill\eject

\singlespace \references \refis{1} S. Coleman, {\it Aspects of Symmetry:
Selected Erice Lectures} (Cambridge University Press; New York,1985), Chapter
7, ``The Uses of Instantons."

\refis{cole} S. Coleman and K. Lee, \journal Phys. Lett. , 221B, 242, 1989.

\refis{cliff} C.P. Burgess and A. Kshirsagar, {\sl Nucl. Phys.}
{\bf B324}, 157 (1989).

\refis{gold} H. Goldstein, {\it Classical Mechanics}, 2nd ed. (Addison-Wesley;
Reading, Mass., 1980), p. 351.

\refis{surf} J.W. York, Jr., \prl 28, 1082, 1972 ; S.W. Hawking, \pr D18, 1747,
1978.

\refis{5} S.B. Giddings and A. Strominger,  \journal Nucl. Phys., B306, 890,
1988.

\refis{6} J. Wess and J. Bagger,  {\it Supersymmetry and
Supergravity} (Princeton University Press; Princeton NJ, 1983).

\refis{West} P. West, {\it Introduction to Supersymmetry and Supergravity}
(World Scientific; Singapore, 1986).

\refis{cremmer} E. Cremmer {\it et al.}, \journal Nucl. Phys., B147, 105, 1979.

\refis{10} Y. Park, M. Srednicki, A. Strominger, \journal Phys. Lett., 244B,
393, 1990.

\refis{k} S.W. Hawking, \pr D37, 904, 1988.

\refis{alpha} S. Coleman, \journal Nucl. Phys., B307, 864, 1988 ;
T. Banks, I. Klebanov, and L. Susskind, \journal Nucl. Phys., B317,
665, 1989 ; R.C. Myers, \journal Nucl. Phys., B323, 225, 1989.

\refis{bagger} J. Bagger, \journal Nucl. Phys., B211, 303, 1983.

\refis{Feyn} R.P. Feynman and A.R. Hibbs, {\it Quantum Mechanics and Path
Integrals} (McGraw-Hill Inc.; New York, 1965).

\refis{andy} S.B. Giddings and A. Strominger, \journal Nucl. Phys., B321, 481,
1989.

\refis{sstev} S.B. Giddings and A. Strominger, \journal Phys. Lett., 230B,
46, 1989.

\refis{big} W. Fischler and L. Susskind, \journal Phys. Lett., 217B, 48, 1989;
S. Coleman and K. Lee, \journal Phys. Lett., 221B, 242, 1989.

\refis{myers} R.C. Myers, \pr D38, 1327, 1988.

\refis{grin} B. Grinstein and J. Maharana,
\journal Nucl. Phys., B333, 160, 1990.

\refis{ab} L.F. Abbott and M.B. Wise, \journal Nucl. Phys., B325, 687, 1989.

\refis{raja}  see for example:
R. Rajaraman, {\it Solitons and Instantons: An Introduction to
Solitons and Instantons in Quantum Field Theory} (North-Holland Publishing
Co.; New York, 1982).

\refis{majic} see for example:
P. van Nieuenhuizen in: B.S. DeWitt and R. Stora, eds.,
{\it Relativity, groups and topology II} (Elsevier Science Publishers;
1984)

\refis{fay} P.D. D'Eath, H.F. Dowker, and D.I. Hughes,
``Supersymmetric Quantum Wormhole States,''
DAMTP preprint R90-23,
presented at 5th Seminar on Quantum Gravity, Moscow, U.S.S.R., May 28
- Jun 1, 1990.

\refis{andyrev} A. Strominger, ``Baby Universes,'' in the proceedings
of the 1988 TASI summer school.

\refis{twilight} see for example:
S.J. Gates, M.T. Grisaru, M. Ro\u cek and W. Siegel, {\it Superspace}
(Benjamin/Cummings Publishing Co., Reading, Massachusetts, 1983).

\refis{ian} I. Affleck, J. Harvey and E. Witten, {\sl Nucl.Phys.}
{\bf B206}, 413 (1982).

\refis{gerry} G. Gilbert, {\sl Nucl.Phys.}
{\bf B328}, 159 (1989).

\refis{kim} K. Lee and S.M. Smirnakis, ``Wormholes Made of Fermions,''
Harvard preprint HUTP-89-A024, May 1989.

\endreferences

\vfill

\head{Figure Captions}

Figure 1: Surfaces of constant Euclidean time in the wormhole geometry, using
radial
($r\-\pm,\ I,\ F$) or rectangular ($t\-\pm,\ i\-\pm,\ f\-\pm$) time slices.

\endit